\documentclass[11pt]{article}

\usepackage{amsmath,amssymb,amsthm}
\usepackage{graphicx}
\usepackage{hyperref}
\usepackage{geometry}
\usepackage[numbers,sort&compress]{natbib}

\title{Geometric and Operational Characterization of Two-Qutrit Entanglement}

\author{Ankita Jana\\Department of Physics, Indian Institute of Technology Jammu, Jammu,\\181221, Jammu \& Kashmir, India}

\date{}
\begin{document}

\maketitle
\begin{abstract}
    
 We investigate the entanglement structure of bipartite two-qutrit pure states
from both geometric and operational perspectives. Using the eigenvalues of the reduced density matrix, we analyze how symmetric polynomials characterize pairwise and genuinely three-level correlations. We show that
the determinant of the coefficient matrix defines a natural, rank-sensitive
geometric invariant that vanishes for all rank-2 states and is nonzero only for
rank-3 entangled states. An explicit analytic constraint relating this
determinant-based invariant to the I-concurrence is derived, thereby defining
the physically accessible region of two-qutrit states in invariant space.
Furthermore, we establish an operational correspondence with three-path
optical interferometry and analyze conditional visibility and predictability in
a qutrit quantum erasure protocol, including the effects of unequal path
transmittances. Numerical demonstrations confirm the analytic results and the
associated complementarity relations. These findings provide a unified
geometric and operational framework for understanding two-qutrit entanglement.
\end{abstract}

\noindent\textbf{Keywords:}
Two-qutrit entanglement; I-concurrence; Determinant invariant;
Symmetric polynomials; Quantum erasure; Multi-path interferometry

\section{Introduction}\label{sec1}

Higher-dimensional quantum systems offer a substantially richer set of correlation
structures compared to two-level systems and therefore play an increasingly important
role in quantum information science \cite{Horodecki2009}. In particular, qutrit-based
architectures have attracted considerable interest due to their potential advantages
in quantum communication, including enhanced security, increased information
capacity, and improved resilience to noise
\cite{Bechmann2000,Kaszlikowski2000,Collins2002}. These features motivate a detailed
investigation of entanglement properties in higher-dimensional bipartite systems.

For bipartite pure states, entanglement is fully determined by the spectral properties
of the reduced density matrix obtained by tracing out one subsystem
\cite{Horodecki2009,Rungta2001}. In two-level systems, this spectral characterization
underlies standard entanglement quantifiers such as concurrence \cite{Rungta2001} and entanglement
of formation. In contrast, higher-dimensional systems exhibit additional correlation
structures that cannot be described solely in terms of two-level coherence. In the
case of two-qutrit states, the presence of three non-negative eigenvalues naturally
allows one to distinguish between correlations arising from pairwise combinations
of levels and those involving all three levels simultaneously
\cite{Rungta2001,Rungta2003}.

Despite extensive progress in the development of entanglement measures and
invariant-based classifications for higher-dimensional systems, a comprehensive
geometric and algebraic understanding of two-qutrit entanglement remains incomplete.
In particular, the interplay between the rank of the reduced density matrix, the
internal structure of the coefficient matrix describing the state, and the connection
between abstract entanglement invariants and experimentally accessible quantities
has not yet been fully clarified within a unified framework
\cite{Albeverio2001,Linden1999}.

Beyond spectral information alone, the algebraic structure of a bipartite quantum
state is encoded in its coefficient matrix. The cofactor structure of this matrix is
directly related to the eigenvectors of the reduced density matrix, indicating a deeper
relationship between matrix algebra, symmetric polynomials of the eigenvalues, and
entanglement properties\cite{Albeverio2001}. Exploring this relationship provides a natural route toward
identifying geometric invariants that distinguish different forms of entanglement in
two-qutrit systems \cite{Albeverio2001,Linden1999}.

While determinant-based invariants and concurrence-type measures have been
investigated independently in the context of higher-dimensional entanglement,
a direct analytic and geometric comparison between a determinant-based invariant
and the I-concurrence for two-qutrit pure states has not been explicitly explored,
to the best of our knowledge
\cite{Rungta2001,Jarvis2014,Osterloh2015}.

In this work, we present a unified geometric and operational characterization of
entanglement in bipartite two-qutrit pure states, building on existing concurrence-based measures and invariant approaches\cite{Rungta2001,Rungta2003,Jarvis2014}. Using symmetric polynomials of the
eigenvalues of the reduced density matrix, we separate contributions associated with
pairwise correlations from those arising from genuinely three-level entanglement.
We show that the determinant of the coefficient matrix defines a natural, rank-sensitive
geometric invariant that vanishes for all rank-2 states and becomes nonzero only for
rank-3 entangled states. An explicit analytic constraint relating this determinant-based
invariant to the I-concurrence is derived, thereby delineating the physically accessible
region of two-qutrit states in invariant space.

Furthermore, we establish an operational interpretation of the geometric invariants
through a correspondence with three-path optical interferometry. Within this
framework, we analyze conditional visibility and predictability in a qutrit quantum
erasure protocol, including the effects of unequal path transmittances. Numerical
simulations are used to validate the analytic results and the associated complementarity
relations. Together, these results provide a unified geometric and operational
perspective on two-qutrit entanglement
\cite{Englert1996,Durr2001,Qureshi2021}.

\section{Theoretical Background}

We consider a bipartite quantum system composed of two qutrits, described within a
$3 \otimes 3$ Hilbert space. Any pure state of this composite system can be expressed
in the computational basis as
\begin{equation}
|\Psi\rangle = \sum_{i,j=1}^{3} C_{ij}\, |i\rangle_A |j\rangle_B ,
\end{equation}
where $C$ is a complex $3 \times 3$ coefficient matrix and
$\{|i\rangle_A\}$ and $\{|j\rangle_B\}$ denote orthonormal bases for subsystems $A$
and $B$, respectively. Throughout this work, the state is assumed to be normalized.

The reduced state of subsystem $A$ is obtained by tracing over subsystem $B$,
leading to the reduced density matrix
\begin{equation}
\rho_A = \mathrm{Tr}_B\!\left(|\Psi\rangle\langle\Psi|\right),
\end{equation}
which can be written explicitly in terms of the coefficient matrix as
\begin{equation}
\rho_A = C C^{\dagger}, \qquad \mathrm{Tr}(\rho_A) = 1 .
\end{equation}
The matrix $\rho_A$ is Hermitian, positive semidefinite, and normalized to unit trace.
Its eigenvalues $\{\lambda_1,\lambda_2,\lambda_3\}$ therefore satisfy
\begin{equation}
\lambda_1 + \lambda_2 + \lambda_3 = 1 .
\end{equation}
For bipartite pure states, the entanglement properties are fully determined by the
eigenvalue spectrum \cite{Horodecki2009,Rungta2001}.

For higher-dimensional bipartite systems, several entanglement measures have been
proposed to generalize the notion of concurrence beyond two-level systems. Among
these, the I-concurrence provides a natural extension applicable to arbitrary finite
dimensions \cite{Rungta2001}. For a two-qutrit pure state, the I-concurrence is defined as
\begin{equation}
C_I = \sqrt{2\left(1 - \mathrm{Tr}\,\rho_A^{\,2}\right)}
    = 2\sqrt{\lambda_1\lambda_2 + \lambda_2\lambda_3 + \lambda_3\lambda_1}
    = 2\sqrt{s_2}.
\end{equation}
Invariant-based approaches and algebraic characterizations of entanglement
in higher-dimensional systems have also been explored in the literature
\cite{Rungta2003,Albeverio2001}.
For a maximally entangled two-qutrit pure state
$\lambda_1=\lambda_2=\lambda_3=\frac{1}{3}$,
the I-concurrence attains its maximum value
\begin{equation}
C_I^{\max} = \frac{2}{\sqrt{3}} .
\end{equation}

This form makes explicit that the I-concurrence depends solely on pairwise products of the eigenvalues and therefore captures correlations associated
with two-level coherences within the qutrit system.

Beyond spectral properties, the algebraic structure of the coefficient matrix
$C$ plays an important role in determining the properties of $\rho_A$. A useful
identity for any $3 \times 3$ matrix $M$ is
\begin{equation}
(M-\lambda I)\,\mathrm{adj}(M-\lambda I)=\det(M-\lambda I)\,I
\end{equation}
where $\mathrm{adj}(\cdot)$ denotes the adjugate matrix \cite{Albeverio2001}. For an eigenvalue
$\lambda = \lambda_i$ of $\rho_A$, the determinant vanishes, implying
\begin{equation}
(\rho_A - \lambda_i I)\,\mathrm{adj}(\rho_A - \lambda_i I) = 0 .
\end{equation}
As a consequence, the columns of the adjugate matrix provide eigenvectors of
$\rho_A$. This establishes a direct connection between the cofactor structure
of the coefficient matrix $C$ and the eigenstructure of the reduced density
matrix\cite{Albeverio2001,Linden1999}.

The eigenvalues of $\rho_A$ also define three elementary symmetric polynomials,
\begin{equation}
s_1 = \lambda_1 + \lambda_2 + \lambda_3 = 1, \qquad
s_2 = \lambda_1\lambda_2 + \lambda_2\lambda_3 + \lambda_3\lambda_1, \qquad
s_3 = \lambda_1\lambda_2\lambda_3 .
\end{equation}
While $s_2$ quantifies pairwise correlations and directly determines the
I-concurrence\cite{Rungta2001,Rungta2003}, the third-order polynomial $s_3$ captures genuinely three-level
correlations that are absent in rank-2 states\cite{Albeverio2001}.This observation motivates the introduction of a rank-sensitive geometric invariant, developed in the following section.

\section{Symmetric Polynomial Geometry and Rank Structure}

The spectrum $\{\lambda_1,\lambda_2,\lambda_3\}$ of the reduced density matrix provides a natural geometric characterization of two-qutrit entanglement through the elementary symmetric polynomials. Since the trace
is fixed by normalization, $s_1=\lambda_1+\lambda_2+\lambda_3=1$, the
entanglement properties are fully determined by the remaining two
independent invariants $s_2$ and $s_3$
\cite{Horodecki2009,Albeverio2001}:
\begin{equation}
s_2 = \lambda_1\lambda_2+\lambda_2\lambda_3+\lambda_3\lambda_1, \qquad
s_3 = \lambda_1\lambda_2\lambda_3 .
\end{equation}

The second-order polynomial $s_2$ quantifies correlations arising from
pairwise combinations of the Schmidt components and directly determines
the I-concurrence, as discussed above \cite{Rungta2001,Rungta2003}. In contrast, the third-order polynomial $s_3$
captures correlations that involve all three levels simultaneously and
therefore represents genuinely three-level entanglement
\cite{Albeverio2001}.

A particularly important consequence of this structure is its sensitivity
to the rank of the reduced density matrix. Rank-2 states are characterized
by the vanishing of one eigenvalue, which implies
\begin{equation}
s_3 = 0 ,
\end{equation}
whereas rank-3 states satisfy $s_3>0$, reflecting the presence of
correlations across all three levels. The polynomial $s_3$ therefore acts
as a sharp indicator of the rank of the entangled state and provides a
natural geometric separation between pairwise and genuinely
three-level correlations in two-qutrit systems
\cite{Albeverio2001,Linden1999}.

This rank sensitivity admits a direct algebraic interpretation in terms of
the coefficient matrix $C$. For a two-qutrit pure state, the determinant of
$C$ is related to the eigenvalues of $\rho_A$ through
\begin{equation}
\det(\rho_A) = |\det C|^2 = \lambda_1 \lambda_2 \lambda_3 .
\end{equation}

Related invariant-based descriptions of bipartite and multipartite quantum
states have been studied from both nonlocal and algebraic perspectives
\cite{Albeverio2001,Linden1999,Jarvis2014}.

The determinant thus vanishes identically for all rank-2 states and is
nonzero only for rank-3 states as expected from invariant consideration \cite{Albeverio2001}. This observation motivates the introduction
of a determinant-based geometric invariant that quantifies genuinely
three-level entanglement \cite{Albeverio2001}.

The symmetric polynomial description therefore provides a clear geometric
picture of two-qutrit entanglement. While $s_2$ determines the strength of
pairwise correlations \cite{Rungta2001,Rungta2003}, $s_3$ distinguishes states
with genuine three-level entanglement from those that effectively reduce to
lower-dimensional subspaces. In the following section, this geometric insight
is used to construct a normalized determinant-based invariant and to derive
an analytic constraint relating it to the I-concurrence.

\section{Determinant-Based Invariant and Analytic Constraint}

The symmetric polynomial structure discussed in the previous section
naturally motivates the definition of a geometric invariant that is
sensitive to genuine three-level correlations. Since the third-order
polynomial $s_3$ vanishes for all rank-2 states and is nonzero only for
rank-3 states, it provides a natural basis for constructing such an
invariant \cite{Albeverio2001}.

We define the normalized determinant-based geometric invariant as
\begin{equation}
G = 3\sqrt{3}\,\sqrt{s_3}
  = 3\sqrt{3}\,\sqrt{\lambda_1 \lambda_2 \lambda_3}.
\end{equation}

The normalization factor $3\sqrt{3}$ is chosen such that $G$ attains its
maximum value unity for maximally entangled two-qutrit pure states, where
$\lambda_1 = \lambda_2 = \lambda_3 = 1/3$ \cite{Horodecki2009}. By
construction, the invariant $G$ vanishes identically for all rank-2 states
and is strictly positive only for rank-3 states, thereby serving as a
faithful indicator of genuine three-level entanglement.

The eigenvalues of the reduced density matrix $\rho_A$ satisfy the
characteristic equation
\begin{equation}
x^3 - x^2 + s_2 x - s_3 = 0,
\end{equation}
whose roots are $\{\lambda_1,\lambda_2,\lambda_3\}$.

Requiring the eigenvalues to be real and non-negative implies that the
discriminant of the cubic polynomial must be non-negative, which leads to
the inequality
\begin{equation}
27 s_3^2 \le 4 s_2^3.
\end{equation}

This inequality follows from the requirement that the discriminant of the
cubic characteristic polynomial be non-negative, ensuring that all
eigenvalues of $\rho_A$ are real and non-negative. Here $s_2$ and $s_3$ are
the elementary symmetric polynomials defined earlier. Physical realizability
of a two-qutrit pure state therefore requires all three roots of the
characteristic equation to be real and non-negative, which imposes a
constraint on the allowed values of $s_2$ and $s_3$ through the
non-negativity of the discriminant \cite{Albeverio2001}.

Expressing the symmetric polynomials in terms of the entanglement measures
$I$-concurrence $C_I$ and the determinant-based invariant $G$ using
$s_2 = C_I^2/4$ and $s_3 = G^2/27$ \cite{Rungta2001,Rungta2003}, the
discriminant condition yields an explicit inequality relating the two
invariants,
\begin{equation}
C_I^6 - C_I^4 - 72 C_I^2 G^2 + 432 G^4 + 64 G^2 \le 0 
\end{equation}

Although both the I-concurrence $C_I$ and the determinant-based invariant $G$ depend solely on the eigenvalues of the reduced density matrix, they capture distinct physical content: $C_I$ quantifies pairwise coherence through second-order eigenvalue correlations, whereas $G$ characterizes genuinely three-level entanglement via the third-order symmetric polynomial.

The resulting inequality delineates the physically admissible region of two-qutrit states in the $(C_I, G)$ plane.
States lying outside this region do not
correspond to valid density matrices and are therefore unphysical. The
boundary of the allowed region is obtained when the inequality is
saturated, which corresponds to spectral degeneracy of the reduced density
matrix, where two or more eigenvalues coincide \cite{Linden1999}.

As a result, the analytic constraint provides a complete geometric characterization of two-qutrit entanglement in terms of the I-concurrence and the determinant-based invariant, consistent with geometric approaches to entanglement classification \cite{Mahanti2023PhysScr}.
The validity of this analytic
constraint is further confirmed by numerical sampling of random two-qutrit
pure states. For all sampled states, the corresponding values of $C_I$ and
$G$ lie strictly within the region defined by the inequality, while the
boundary is saturated only in cases of spectral degeneracy of the reduced
density matrix.

The inequality further highlights the complementary roles of the two invariants. 
While the I-concurrence captures correlations associated with pairwise coherences \cite{Rungta2001,Rungta2003, Roy2022PRA}, 
the invariant $G$ quantifies genuinely three-level entanglement. 
Together, they provide a minimal and complete description of the entanglement geometry of two-qutrit pure states.

\section{Optical Interpretation and Quantum Erasure}

The geometric invariants introduced above admit a natural operational
interpretation in the context of multi-path optical interferometry. In
particular, a two-qutrit system can be mapped onto a three-path
interferometer, where each qutrit basis state corresponds to a distinct
propagation path. Within this framework, the reduced density matrix
elements acquire direct physical meaning in terms of path populations and
interference visibilities \cite{Englert1996,Durr2001,Qureshi2021}.

For a three-path interferometer with relative phase shifts
$\{\phi_1,\phi_2,\phi_3\}$, the detected intensity can be written as
\begin{equation}
I(\phi_1,\phi_2,\phi_3)
= \sum_i \rho_{ii}
+ 2 \Re \sum_{i<j} e^{i(\phi_i-\phi_j)} \rho_{ij}.
\end{equation}

Quantitative formulations of wave–particle duality have been extended to
multi-path and multibeam interferometers, which are particularly relevant
for higher-dimensional systems \cite{Jakob2010,Qureshi2017}.

where the diagonal elements $\rho_{ii}$ correspond to individual path
intensities and the off-diagonal elements $\rho_{ij}$ encode pairwise
interference contributions. The exponential phase factors determine the
positions of the interference fringes, while the real part gives rise to
observable oscillations. In this way, the coherence structure of the reduced
density matrix is directly reflected in the interference pattern
\cite{Englert1996,Durr2001,Jakob2010}.

To analyze the effects of which-path information, we consider a marked
three-path state of the form
\begin{equation}
|\Psi\rangle = \sum_{i=1}^{3} c_i |\pi_i\rangle |\sigma_i\rangle ,
\end{equation}
where $|\pi_i\rangle$ denote the path states and $|\sigma_i\rangle$ represent
the corresponding marker states \cite{Englert1996,Greenberger1988}.

In a realistic interferometric setup, each path may be subject to an
amplitude transmittance $t_i \ge 0$, yielding the effective state
\begin{equation}
|\Psi\rangle = \sum_{i=1}^{3} \sqrt{t_i}\, c_i |\pi_i\rangle |\sigma_i\rangle .
\end{equation}

In an erasure protocol, the marker subsystem is projected onto a
chosen erasure state $|e\rangle$ \cite{Scully1991,Schwindt1999}.
 Defining the
overlap amplitudes $\alpha_i = \langle e | \sigma_i \rangle$ and the
corresponding intensities $\tau_i = |\alpha_i|^2$, the post-selected
(unnormalized) path state becomes
\begin{equation}
|\psi_e\rangle = \sum_{i=1}^{3} \sqrt{t_i}\, c_i \alpha_i |\pi_i\rangle .
\end{equation}

The success probability of the erasure process is given by
\begin{equation}
P_e = \sum_{i=1}^{3} |c_i|^2 t_i \tau_i ,
\end{equation}
and the normalized conditional path state follows straightforwardly.

The conditional predictability after erasure is determined by the path
populations
\begin{equation}
p_i^{(\mathrm{cond})}
= \frac{|c_i|^2 t_i \tau_i}{\sum_k |c_k|^2 t_k \tau_k},
\end{equation}
which satisfy $\sum_i p_i^{(\mathrm{cond})} = 1$. A suitable normalized
measure of qutrit predictability is then defined as\cite{Jakob2010,Qureshi2021}
\begin{equation}
P_{\mathrm{cond}}
= \sqrt{\frac{3}{2}\sum_{i=1}^{3}
\left(p_i^{(\mathrm{cond})} - \frac{1}{3}\right)^2 } ,
\end{equation}
generalizing the standard two-path predictability to three-path
interferometry \cite{Durr2001,Jakob2010,Qureshi2017}.

The recovered interference after erasure is quantified by the conditional
visibility. Summing the contributions from all unordered path pairs yields
\begin{equation}
V_{\mathrm{cond}}
= \frac{\sum_{i<j} 2 |c_i c_j| \sqrt{t_i t_j}\sqrt{\tau_i \tau_j}}
{\sum_k |c_k|^2 t_k \tau_k},
\end{equation}
which captures the restoration of interference due to the erasure of
which-path information \cite{Englert1996,Qureshi2021,Shah2017Pramana}.

Several complementary frameworks, ranging from coherence-based descriptions to delayed-choice and optimal tradeoff approaches, have been explored in the literature
\cite{Baumgratz2014,Kim2000,Jaeger1995,Coles2014,DuerrRempe2000,Bagan2016}.

These expressions reduce to the corresponding ideal forms in the limit of unit
transmittance and perfect erasure.

The determinant-based geometric invariant also acquires a simple operational
modification in the presence of unequal transmittances
\cite{Englert1996,Durr2001,Jakob2010}.
Denoting by $G$ the ideal invariant, the transmittance-weighted form reads
\cite{Qian2018,Coles2014}

\begin{equation}
G_T = 3\sqrt{3}\,\sqrt{t_1 t_2 t_3}\sqrt{\lambda_1\lambda_2\lambda_3}.
\end{equation}

The conditional quantities satisfy the extended complementarity relation
\cite{Coles2014,Bera2015}
\begin{equation}
P_{\mathrm{cond}}^2 + V_{\mathrm{cond}}^2 + G_T^2 \le 1 .
\end{equation}

The role of entanglement in constraining wave–particle duality relations has been
explicitly analyzed in interferometric settings \cite{Jaeger1995,Coles2014,DuerrRempe2000}.
This inequality expresses a three-way tradeoff between which-path information,
recovered two-path coherence, and genuine three-level geometric coherence
\cite{Qureshi2021,Bera2015,Qian2018}.

The relevance of multi-level coherence and entanglement extends to practical
quantum information protocols implemented on current quantum hardware
\cite{Joy2020}.

\section{Numerical Demonstrations}

In this section, we present numerical simulations to illustrate the analytic
constraints and operational relations derived in the previous sections.
The numerical results are intended to provide explicit confirmation of the
geometric structure of two-qutrit entanglement and its operational manifestation
in quantum erasure protocols \cite{Englert1996,Qureshi2021,Bera2015}.

Random two-qutrit pure states were generated by sampling complex coefficient
matrices $C$ subject to normalization. For each realization, the reduced density
matrix $\rho_A = C C^\dagger$ was constructed and its eigenvalues were computed.
The corresponding I-concurrence $C_I$ and determinant-based invariant $G$ were
then evaluated using the definitions introduced earlier
\cite{Rungta2001,Rungta2003}. Only physically valid states satisfying the analytic
constraint of Section~4 were retained.

\begin{figure}[t]
\centering
\includegraphics[width=0.6\linewidth]{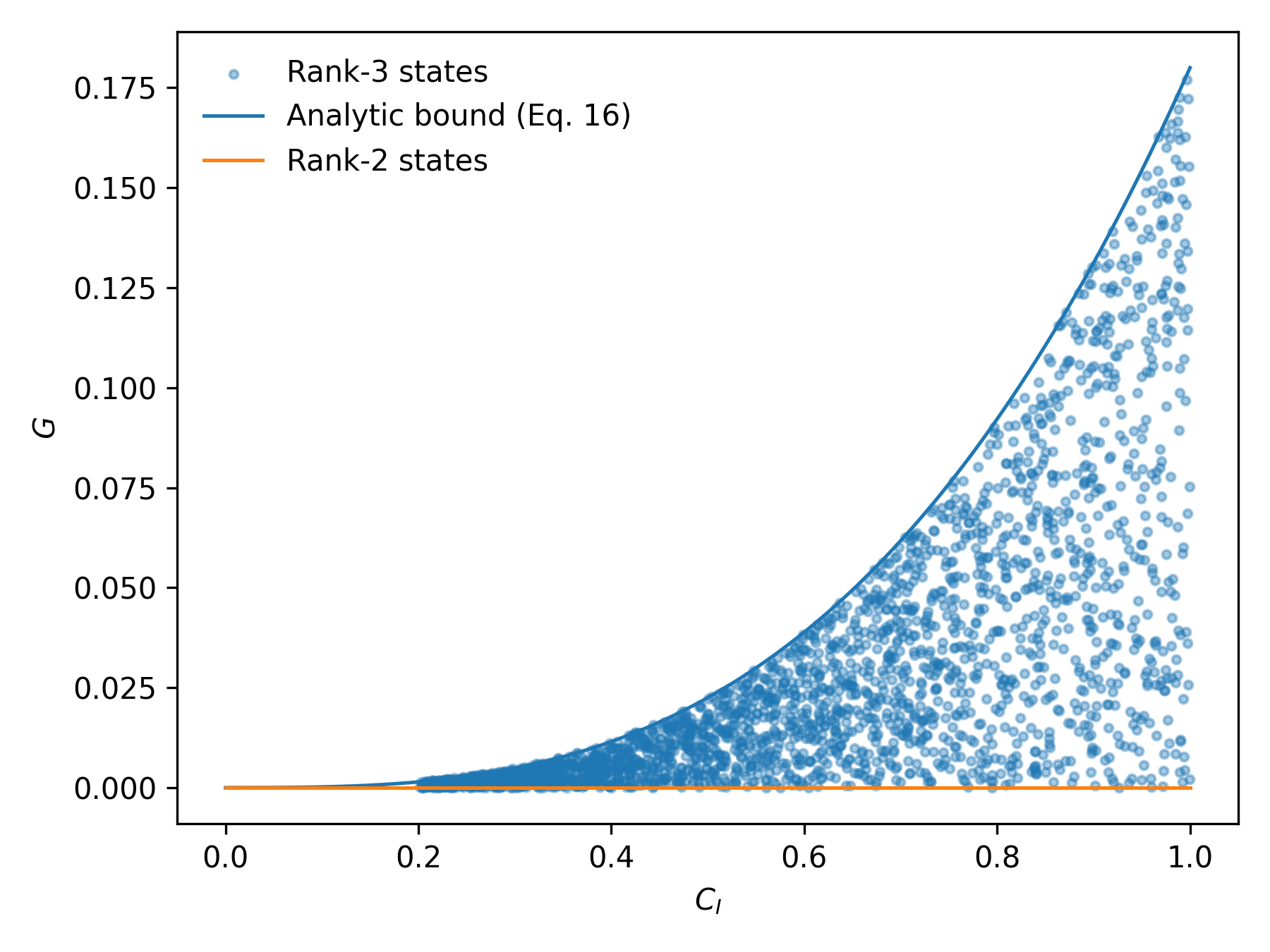}
\caption{Scatter plot of randomly generated two-qutrit pure states in the $(C_I, G)$ plane. Rank-2 states lie on the line $G = 0$, while rank-3 states populate the interior of the physically allowed region. The solid curve denotes the analytic boundary obtained by saturating Eq.~(16).}

\label{fig:CI_G_scatter}
\end{figure}

Figure~\ref{fig:CI_G_scatter} shows the distribution of two-qutrit states in the $(C_I,G)$ plane.
Rank-2 states lie entirely on the line $G=0$, while rank-3 states populate the interior of the
allowed region bounded by the analytic inequality corresponding to spectral degeneracy of the reduced density matrix. The numerical
points are seen to lie strictly within this boundary, confirming the validity of the analytic constraint.

\begin{figure}[t]
\centering
\includegraphics[width=0.6\textwidth]{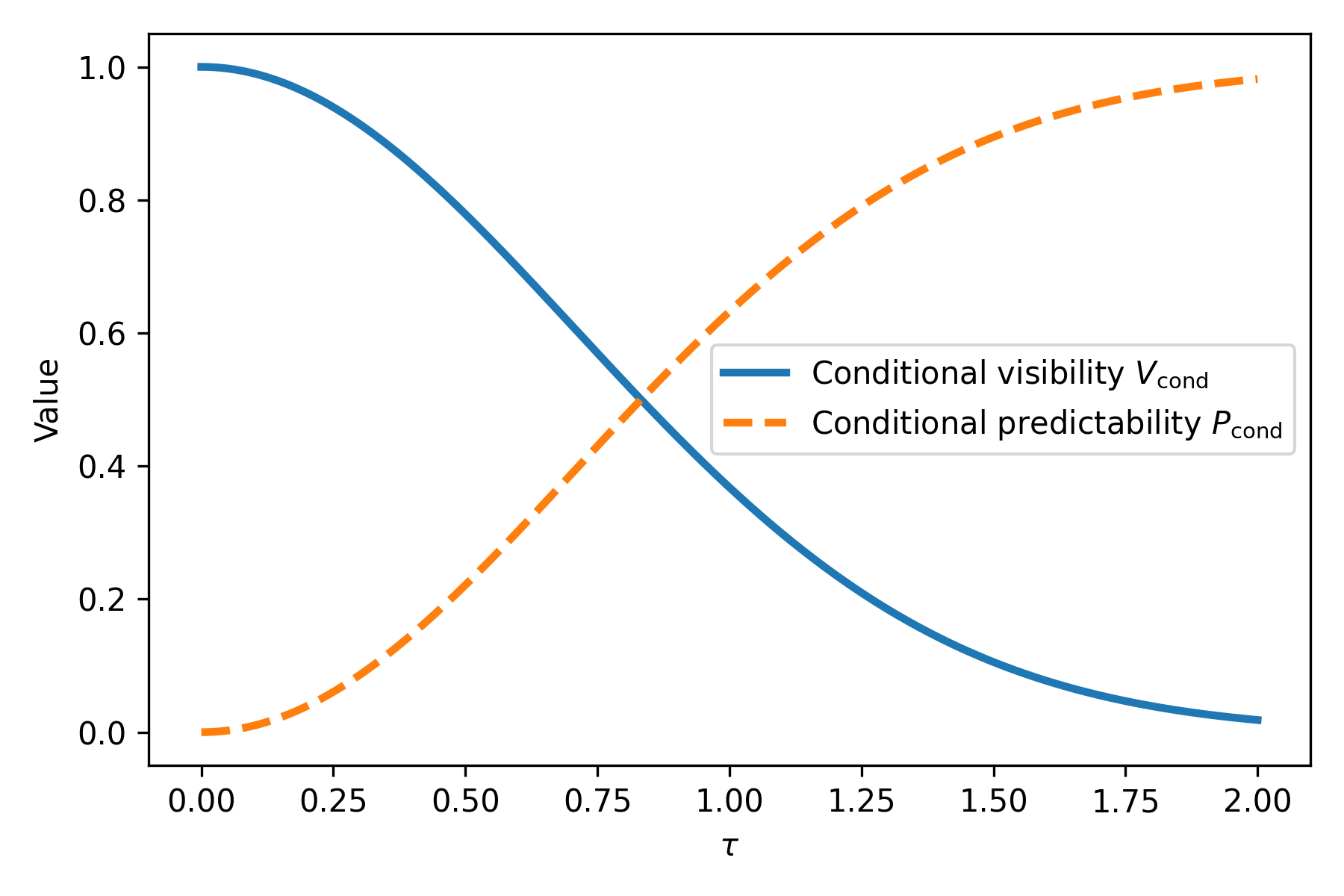}
\caption{Conditional visibility $V_{\mathrm{cond}}$ and conditional predictability $P_{\mathrm{cond}}$ as functions of the erasure parameter $\tau$. The tradeoff between the two quantities demonstrates wave--particle complementarity in the qutrit quantum erasure scheme.}

\label{fig:VP}
\end{figure}

We next illustrate the behavior of conditional visibility and predictability in
the two-qutrit quantum erasure protocol.Figure~\ref{fig:VP} displays the variation of the
conditional visibility $V_{\mathrm{cond}}$ and predictability $P_{\mathrm{cond}}$
as functions of the transmittance parameter $\tau$, demonstrating the expected
complementary behavior \cite{Englert1996,Durr2001,Jakob2010,Qian2018}.

The numerical results are fully consistent with the extended
complementarity relation derived in Section~5. Deviations from saturation
arise from nonidealities such as unequal transmittances or partial erasure,
while the overall tradeoff structure remains intact. These simulations
therefore provide direct numerical support for the unified geometric and
operational framework developed in this work.

\section{Conclusion}

In this work, we have presented a unified geometric and operational description
of entanglement in bipartite two-qutrit pure states. By analyzing the eigenvalue
structure of the reduced density matrix through symmetric polynomials, we have
clearly separated pairwise correlations from genuinely three-level
correlations \cite{Horodecki2009,Rungta2001,Rungta2003}. This approach provides a transparent geometric framework for
understanding higher-dimensional entanglement.

We have shown that the determinant of the coefficient matrix defines a natural,
rank-sensitive geometric invariant that vanishes for all rank-2 states and is
nonzero only for genuine rank-3 entanglement. An explicit analytic constraint
relating this determinant-based invariant to the I-concurrence was derived,
thereby identifying the physically accessible region of two-qutrit states in
invariant space. This constraint offers a complete geometric characterization
of two-qutrit entanglement in terms of two complementary quantities.

An operational interpretation of the geometric invariants was established
through a mapping to three-path optical interferometry. Within this framework,
we derived expressions for conditional visibility and predictability in a
qutrit quantum erasure protocol, including the effects of unequal path
transmittances. The resulting extended complementarity relation reveals a
three-way tradeoff between which-path information, recovered two-path
coherence, and genuine three-level geometric coherence\cite{Englert1996,Durr2001,Jakob2010,Qureshi2021}.

The combination of analytic results and numerical demonstrations confirms the
consistency and robustness of the proposed framework. The geometric and
operational connections developed here provide new insight into the structure
of higher-dimensional entanglement and its observable manifestations. Future
work may explore extensions to mixed states, higher-dimensional systems, or
experimental implementations in multi-path interferometric platforms.
The present results are consistent with broader efforts to understand
the role of coherence and entanglement in quantum information theory\cite{Bera2015,Baumgratz2014}.

\bibliographystyle{unsrt}

\bibliography{references}

\end{document}